\documentclass[letterpaper,onecolumn,prl,showpacs]{revtex4-1}
\usepackage[latin9]{inputenc}
\usepackage{amsmath}
\usepackage{amssymb}
\usepackage{graphicx}
\usepackage{geometry}                			
\usepackage{amssymb}
\usepackage{amsmath}
\usepackage{graphicx}
\usepackage{caption}
\usepackage{subcaption}
\usepackage{dsfont}

\usepackage{epsfig}\usepackage{color}\usepackage{booktabs}

\makeatother

\begin{document}

\title{Cultural Evolution as a  Non-Stationary Stochastic Process}
\author{Arwen E. Nicholson  and Paolo Sibani}
\affiliation{FKF, University of Southern Denmark DK5230 Odense M, Denmark}

\begin{abstract}
We present an individual  based   model of cultural evolution,  where
interacting agents  are coded by binary strings standing  for  
strategies for action,  blueprints for products or attitudes and beliefs.
The model is patterned on an established model of 
biological evolution, the  Tangled Nature Model  (TNM), where a  `tangle'  of interactions  
between agents determines their  reproductive 
success. In addition, our agents also have the ability  to copy part of each other's strategy,
a feature inspired by the Axelrod model of cultural diversity. Unlike the latter, but similarly
to the TNM, the model dynamics goes through a series of metastable stages of increasing length,
each characterized by mutually enforcing cultural patterns. These patterns are abruptly replaced
by other patterns characteristic  of  the next metastable period.
We analyze the time dependence of the population and diversity in the system,
show how different cultures are formed and  merge, and how their  survival probability
 lacks, in the model,  a finite average life-time.
Finally, we use historical data on the number of car manufacturers 
after the introduction of the automobile to the  market, to argue that our
 model can qualitatively  reproduce the flurry of 
cultural activity which follows a disruptive innovation.
\end{abstract}
\maketitle

\section{Introduction} 
Computer models of biological~\cite{Drossel01}  and 
 social~\cite{Castellano09}  evolution
 often involve networks of   interacting agents
 with a  stochastic dynamics able to reach 
a stationary or steady state, which   can then  be given 
a  biological or cultural
 iterpretation~\cite{Rikvold03,Axelrod97}.  The Axelrod model~\cite{Axelrod97} is a
 case in point, 
 where agents  placed on a grid exchange traits with 
their  neighbors with a probability proportional to the number of 
 traits already shared.
Making  use of  two social mechanisms: social influence - the tendency of interacting individuals 
 to become more similar, and homophily - the tendency for individuals to associate with similar others,
this model  quickly reaches an equilibrium state, where interacting agents within spatially localized areas
 have identical traits, possibly differing   from those   in   neighboring areas. 
 This outcome, which   is interpreted in terms of cultural  diversity, 
 strongly  differs  from the \emph{punctuated equilibrium dynamics} of  biological macro-evolution~\cite{Gould93,Gould02}, where 
 equilibrium is  never reached, but is replaced by a series of increasingly  long-lived metastable states.
  Similarly, human  history  is  subdivided into
  successive  metastable periods, each 
 identified by  the   technologies mastered in the period, from the use of fire,
 through  stone and metal  tools,  and up to our current modern technologies.
 Cultural and biological evolution are to some degree intertwined processes, 
since  the   abilities  to communicate, use tools and form societal structures~\cite{Diamond92},
all aspects   of    human culture, have  co-determined the course of early human evolution~\cite{Leland00}.
It seems  therefore justified  to adapt biologically motivated computer models  to study 
 cultural evolution on the computer.
 
The Tangled Nature Model~\cite{Christensen02,Hall02,Anderson04,Lawson06} 
 (TNM) of biological evolution features a sequence of  Quasi Evolutionary  Stable States (QESS)
 of increasing duration,
 during which  aggregated quantities only  vary around fixed average values. The transitions
 between consecutive  QESS  are rapid and turbulent events, called \emph{quakes},
 which  entail considerable rearrangements of the network structure~\cite{Becker14}.
 The model presently introduced  modifies    the 
 TNM by adding  one   feature 
 inspired by the Axelrod model. Even though the changes introduced  hardly
 affect the basic dynamical mechanisms of the TNM, 
we nevertheless    for brevity refer to the resulting   TNM version as the Tangled Axelrod Model, or TAM.
 
Postponing  technical   details   to  the next  section,  the properties    of the
 TNM's and  TAM's   
   are first  summarised  below:
 The  interacting agents of the TNM are represented by binary strings
 which  can   be interpreted   in biological terms as genomes. We note that
 the TNM does not discriminate  between genotype  and phenotype and that 
 the genome can also be interpreted as the carrier 
 of   cultural features, i.e. blueprints or strategies for action. The latter 
 interpretation is the one carried over to the TAM, even though we keep the
 word `genome'  to refer to the string bits characterising TAM agents.
 
TNM agents  reproduce asexually and in  error prone fashion at a rate   which depends on  the  `tangle'  of interactions
 connecting  them to each other, with positive, or mutualistic, interactions leading  to a higher reproduction rate.
 Since  extant agents draw resources from a shared and finite   pool, they all  
 have an   indirect, global and  negative effect  on each  other's reproductive success.
 Importantly, interactions between  two TNM  individuals are fully determined by 
a random but  fixed function of their genomes.  
A TNM agent can  thus be labeled  in two different but equivalent 
ways, either using the genome itself   or  using the set  of all
interactions that the genome generates together with other genomes.
 In the following,  agents connected by
non-zero interactions will be called \emph{acquaintances}.
Agent  removals happen  at a  constant rate and independently of interactions.

 Each TAM agent's  genome has  two parts  of equal size. 
The first part, called \emph{interaction genome}
 determines the interactions, just like 
in the TNM.  The second part, called \emph{cultural genome } or
\emph{strategy} is available for other agents to copy in
full or in part.
Sets  of agents with the same interaction genomes constitute a 
  \emph{family}, and sets  of agents with the same strategies 
 constitute  a  \emph{culture}.
  
in the
TAM,  subsets of randomly grouped  agents, termed \emph{neighbors},
are able to copy parts of each other's strategy.
When, say,  agent  `a',  copies part of  the strategy 
of a neighbor `b',  a new agent, `c' is produced. The latter  
 inherits the  family and the neighbors of `a', and a mix of   
  `a's  and `b's strategies.
Unlike  the Axelrod model, there is no spatial grid in the TAM
 and   neighbors  have no spatial relation to one another. 
 Our  way of distributing the interactions makes sense  
 with  internet and mass media connecting people worldwide.
In  the Axelrod model, the probability of  one agent inducing
a change   in the   traits of another   is proportional to the pre-existing
overlap of traits between the two. In our model, overlap is not available
information for agents decisions, and  strategies are copied 
with a probability given by  their relative frequency in the population,
i.e. popular strategies are more likely to be copied.

Following the nomenclature developed in Ref.~\cite{Becker14}, to which we refer for 
an in-depth  discussion of TNM dynamics, extant species in the TNM 
are divided into a  \emph{core} and a \emph{cloud}. Core species have, by definition,
 a population 
exceeding  $5\% $ of  the most populous species. As it turns out, they have   mutualistic  interactions
with one another  and 
together constitute the metastable \emph{core} characterizing  each QESS in a TNM trajectory.
Cloud  species are sparsely and intermittently populated, mainly by an influx of mutants
from the core. The interactions between cloud species are distributed in the same
way as the interactions between unsorted species.
Core and cloud tougher make up an \emph{ecosystem}.
The core and cloud  definitions are here extended to TAM families and cultures.  
`Culture' can of course  have different interpretations, e.g.
production  technologies, languages or even fashion. 
Only core families 
are those sufficiently populous and stable to deserve the name of  cultures, but the term
is  used for simplicity    for  all families, even though the `cultures' cloud families can 
carry  are only tantamount to random noise. 

The generic properties of the TAM are as follows:
Starting out from a single    family with a single culture,
core species with different cultures soon appear,
their number growing slowly  as a function of time.
An established  culture can either disappear abruptly in a \emph{quake}  together  with all its proponents
or more slowly  as   more popular strategies
get copied and eventually  take over.
The probability that a culture extant at time $t_{\rm w}$ 
remains  so at time $t>t_{\rm w}$ depends on both $t_{\rm w}$ and $t$,
and decays with $t$ in a power-law like fashion. Cultures in the TAM lack
a finite average life-time, which translates into an expected  large variation of
the duration of actual cultures.
The  statistical properties of the model follow from 
 a minimum of assumptions. 
 In particular, our agents' decisions are stochastic rather then 
 rational  and are  based on  a knowledge of
the situation which is  limited in both  time and space. Due to these minimalistic assumptions,
we  suggest  that the TAM might  serve 
 as a generic null model of cultural evolution.
 
 \newpage
\subsection*{Summary of nomenclature}
This section is  a short summary of the nomenclature used in this work, highlighting
 differences and similarities between the TNM and the TAM. It makes reference to 
 details explained in the next Section.

\begin{description}
\item[Genome] In the TNM the genome is a string of $N$ bits with an obvious biological interpretation. In the TAM
the same word is used for historical reasons to describe a string of $2N$ bits.
The first $N$ bits constitute the interaction genome and 
determine all the interactions with other agents.
The last $N$ bits can be copied by other agents, are understood as 
 a \emph{blueprint} for action and are
called \emph{strategy}. There is no biological interpretation.
 \item[Strategy]   In the TAM context, denotes the part of the  \emph{genome}
 that other agents can copy see above.
\item[Interactions] Interactions determine the reproductive success of both TNM and TAM agents.
In the TNM, interactions between individuals are uniquely determined by their genomes. In the TAM they
are determined  in a similar fashion by the interaction part of the genome.
\item[Mutations] In the TNM mutation can hit anywhere in the genome. In the TAM they only affect the strategy part of the genome.
The interaction part is affected indirectly, as described in the next section.
\item[Species] Set of TNM individuals with the same genome.
\item[Family] Group  of TAM individuals each endowed with the same set of interactions. Not used in a TNM context, where 
the term 
corresponds  to a \emph{species}.
\item[Culture] Group  of TAM individuals each endowed with the same strategy. Not used in a TNM context.]
\item[Core species] TNM species with at least 5\% of the most populous species.
\item[Core culture/family] TAM culture/family with at least 5\% of the most populous culture/family.
\item[Trait] The part of the genome/strategy of a TAM individual which can be copied in a single copying attempt. Not used for TNM.
\item[Neighbor] Each TAM agent can  exchange traits with its 	neighbors. 
\item[Acquaintance] Both TNM and TAM agents have non-zero interactions with their acquaintances. 
\end{description}

\section{TAM implementation}
Our model's elementary dynamical variables are two binary strings of length $K$,
which together characterise an individual.
The first string, the   \emph{interaction genome} is a 
point of the $K$ dimensional hypercube, which 
  is populated by a  \emph{family} of individuals with the same
  interaction genome.
  A similar grouping can be done using the 
  $K$ bit string, called  \emph{strategy}. A strategy is a point in a distinct $K$ dimensional hypercube,
  populated by individuals  sharing the same  \emph{culture}.
    Simulation time is given in \emph{generations}, each comprising the number of updates needed to remove the  extant population.
Initially, the  population  is $N$ and a generation comprises
$N/p_{\rm kill}$  updates. Later,  the   generation length is  computed similarly, but
using  the population present  at the end of the preceding generation.
 Unlike the TNM, and due to the copying and mutation mechanism introduced,
the interaction genome does not uniquely identify  individuals. A second $K$ bit string,
the \emph{strategy} is attached to each individual and is subject to copying and mutation.
 
Neglect  first the effect of copying and
consider two agents, $a$ and $b$, still having the interaction genome and strategy  they inherited from their parent,
possibly modified by  point mutations. To generate the interaction between the two, three fixed arrays of length $2^K$, $\mathbf I$,
$\mathbf F_1$ and $\mathbf F_2$ are utilized. The first contains ones with probability $p_{\rm acquaintances}=1/4$ 
and zeros otherwise,
and the other two contain random numbers drawn from a Gaussian distribution with  zero mean and unit variance.
The interaction genomes of $a$ and $b$ are first XOR'red to produce a new binary string $c$. Now reading binary strings as integer labels  when needed, 
the coupling $J_{ab}$ is zero if and only if $I(c)=0$. Otherwise,  $J_{ab}=G F_1(c)F_2(b)$, where $G$ is a constant.
Agents can copy parts of each other's strategy  if they are \emph{neighbours}. Our agents  $a$ and $b$ are neighbours if and only if $L(c)=1$,
where the string $c$ is obtained as just described, and where $\mathbf L$ is an array containing  ones with probability $p_{\rm neighbors}=1/4$ and zeros otherwise.

Reproduction probabilities in the TAM  are calculated 
as in the TNM:  Let $\mathcal{S}$ denote the ecosystem, $N_b(t)$ denote the population size of species $b$, and $N(t) = \sum_\mathcal{S} N_b(t)$ be the total population size.
  An individual of type $a$ is chosen as candidate for reproduction  with probability $n_a = N_a/N$, and successfully reproduces with probability %
$p_{\rm off}(a) = 1/(1 + e^{-H_a})$, where %
\begin{equation}
 H_a(t) = -\mu N(t)  + \sum_b j_{ab}(t),
\label{eq:Pfunc}
\end{equation}%
and where
\begin{equation}
j_{ab} =\frac{N_b}{N} J_{ab}= J_{ab} n_b
\label{dens_w_c}
\end{equation}%
is a density weighted coupling.
In Eq.~\eqref{eq:Pfunc}, $\mu$ is a positive constant which limits the size of the ecosystem. 
At each successful reproduction step, a point mutation in the strategy of the offspring occurs  with
probability of mutation per bit by $p_{\rm mut}$. Parent and offspring strategies then differ by $k$ bits
 with  probability   Bin$(k;K,p_{\rm mut})$, the binomial distribution. Natural death occurs with  probability $p_{\rm kill}$.
 A copying move is performed with probability $p_{\rm copy}$, and 
 an individual of type $a$ is chosen to do the copying 
  with probability $n_a = N_a/N$. A second individual $b$ is then   picked from a list of its neighbors 
 with uniform probability.
Strategies are  partitioned in $k$ segments of equal  length, called \emph{traits}. 
 Agent $a$ copies 
  $k$ randomly chosen   traits from $b$ with probability 
    Bin$(k;K/k,P(a,b))$, where 
 \begin{equation}
 P(a,b) =\frac{C(b)}{C(a)+C(b)}
 \label{copyingprob}
 \end{equation}
and  where $C(x)$ is the number of individuals sharing the same culture as 
 agent $x$, independently of these agents interaction genomes.
 The outcome  of the process is a new agent $d$ with  the same interaction genome, acquaintances and neighbors 
 as parent $a$, the parent doing the copying.
 Agents with a copied strategy  and their  progeny     maintain  the interactions and neighbors of
  their original parent until a strategy mutation occurs.  Then, 
 the interaction genome is set  equal to the newly mutated strategy and new interactions are  generated 
 according to the prescription  described above.

\begin{figure}
	\includegraphics[width=\linewidth]{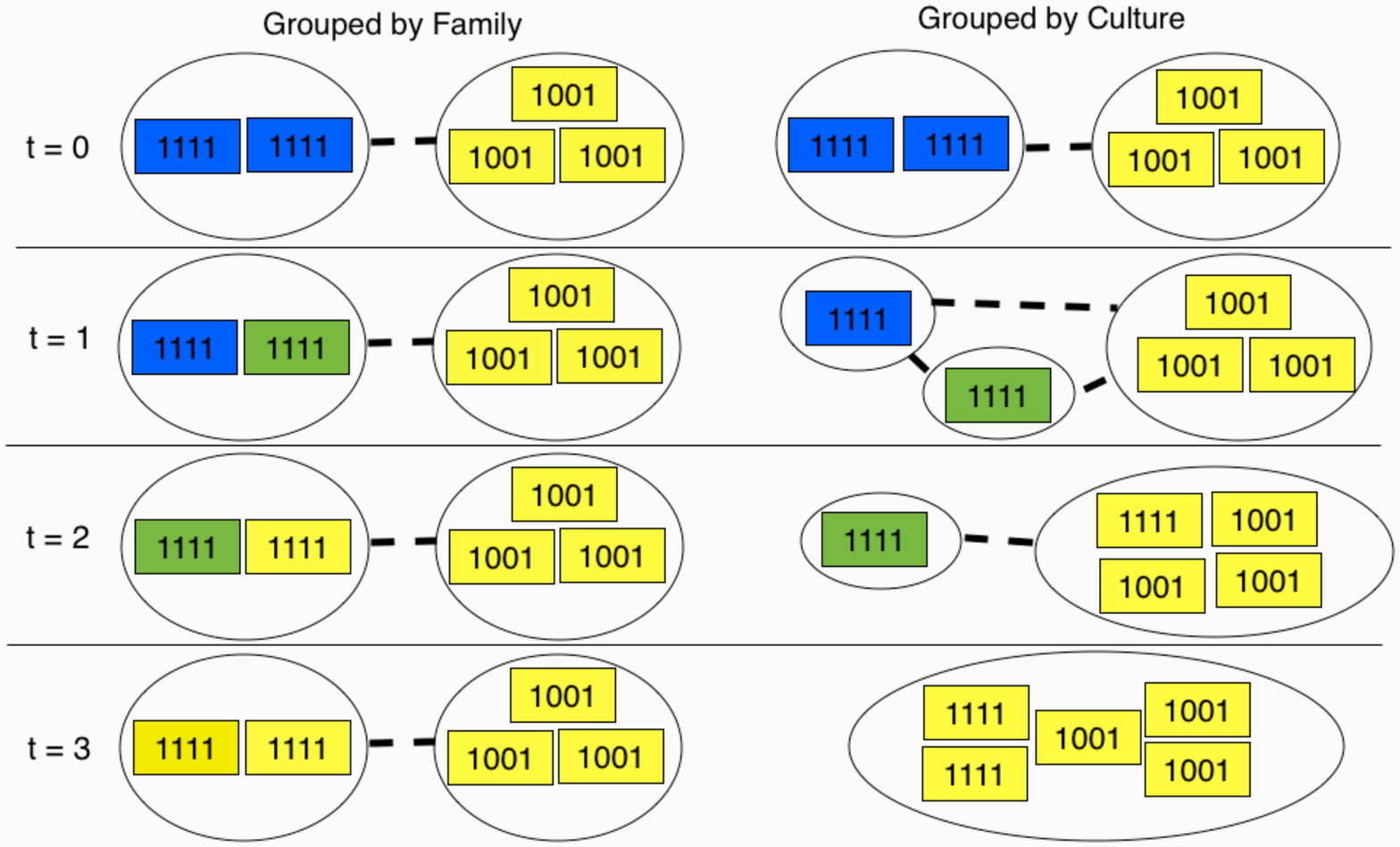}  
\caption{  Illustration of the TAM copying
mechanisms. For details see main text below.
	}
 \label{illustration}
\end{figure}

The TAM copying  mechanism is illustrated in Fig.~\ref{illustration}.
In the figure,  the interaction genome of individuals is a four bit string. Their
likewise four bit strategy is omitted and   replaced  by a color code for visual
clarity. The dashed lines represent neighbour relations and individuals are grouped either by family
or by culture.  At time $t=1$,  a blue individual copies part of the strategy of a yellow
individual, resulting in a green individual with the same interaction code, $1111$, as its copying parent.
At time $t=2$ the green individual copies the  yellow strategy, resulting
in a new yellow individual, with interaction string $1111$. 
At time $t=4$ the green culture has disappeared completely, leaving the yellow culture behind.
Note that the individuals of this culture have different interaction genomes and hence different 
interactions.

 The TAM mutation mechanism indirectly generating  all new 
  interaction patterns only affect s individuals with  a copied strategy
  and 
leads  to  an  exploration of configuration space which is faster   than 
 is the case in the TNM model.   As a consequence, when
 $p_{\rm copy} \ge p_{\rm mut}$ the systems  
 become more often  unstable and a  number of simulations ending 
  in extinctions is  significantly higher than in the TNM case. If the (still rare) cases where the system is heading to 
  extinction, the last individual is not killed but replaced by $500$ of its clones.
 This is similar to the usual starting conditions,
   except that the individuals are not placed on a random point in  cultural space,
   but keep the position already attained. 

 To summarize, unless otherwise stated, the following parameters were used in the simulations:
 \begin{itemize}
\item  initial population contains  $500$ identical agents.
 \item   environmental harshness $\mu  = 0.1$.
\item  interaction coupling strength $G = 100$.
\item  strategy  and interaction genome have each  length $L = 20$.
\item each trait has length one and the  two possible states $\pm 1$.
\item probability of  being connected as acquaintances: $p_{\rm acquaintances} = 0.25$.
\item probability as being connected as neighbours: $p_{\rm neighbors} = 0.25$.
 \item probability of death: $p_{\rm death} = 0.2$.
\item probability of strategy mutation : $p_{\rm mut} = 0.01$.
\item  probability of trait copy attempt from a  neighbor: $p_{\rm copy} = 0.01$.
 \end{itemize}
\section{Results}
\begin{figure}
        \centering
        \begin{subfigure}[b]{0.46\textwidth}
                \centering
                \includegraphics[width=\textwidth]{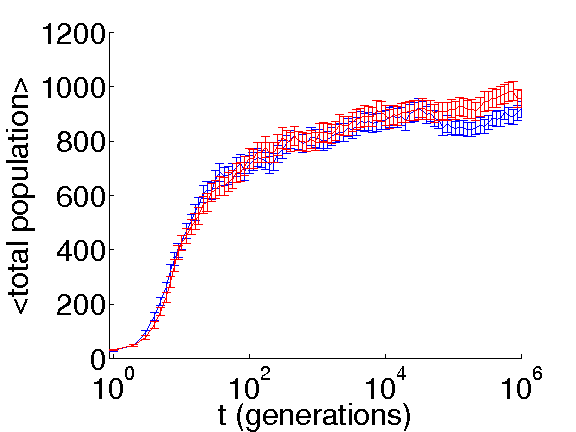}
                \label{TAS1}
        \end{subfigure}%
        ~ 
        \begin{subfigure}[b]{0.46\textwidth}
                \centering
                \includegraphics[width=\textwidth]{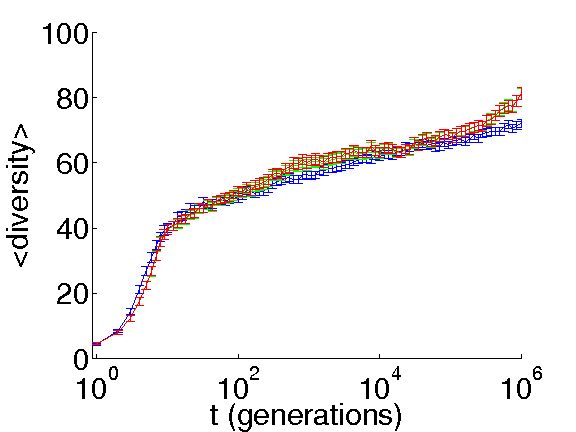}
                \label{TAS2}
        \end{subfigure}
        ~ 
        \begin{subfigure}[b]{0.46\textwidth}
                \centering
                \includegraphics[width=\textwidth]{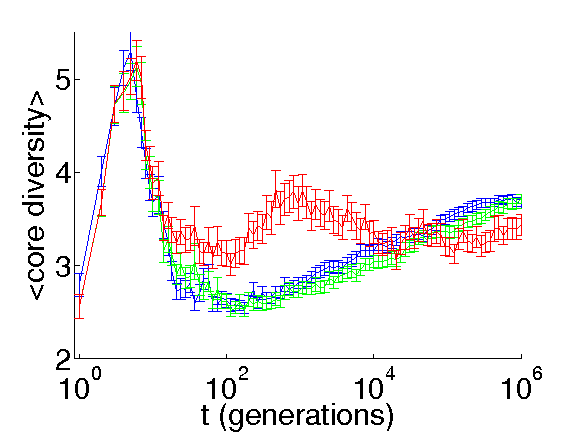}
                \label{TAS3}
        \end{subfigure}
          ~ 
        \begin{subfigure}[b]{0.46\textwidth}
                \centering
                \includegraphics[width=\textwidth]{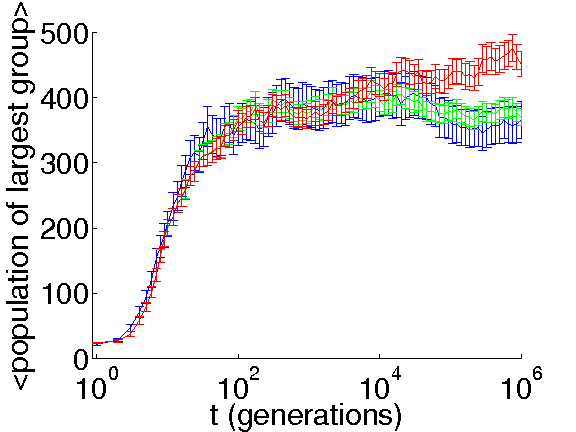}
                \label{TAS4}
        \end{subfigure}
        \caption{Blue refers to TNM data, red to TAM data grouped by culture, and green to TAM data grouped by family.
        Top left and right: Total population (independent of grouping)  and total diversity vs. time.
        Bottom left and right: Core diversity and population of the largest group vs. time. Note the
        logarithmic abscissa. All averages are taken 
        over 200 trajectories. }
        \label{macro_data}
\end{figure}
Each panel of figure~\ref{macro_data} displays, with 
 one-$\sigma$ error bars,  three different 
 time series, each  the outcome of averaging  200 independent trajectories. 
The nearly undistinguishable blue  and green data
  pertain to  TNM and  TAM families, while the red data pertain to  
 TAM cultures.
 The total population,  which does not depend on the way in which
 individuals are grouped, is plotted vs. time 
 in the   uppermost left panel.  We see a small difference between the TNM (blue) and
 TAM (red) data, both growing logarithmically in time after a short initial
 transient. A similar behavior is observed for the diversity, which is
 plotted in the uppermost right panel.  The diversity of e.g.  families is the number
 of different families extant at a certain time.
 The lower left panel shows the family and cultural diversity of the  core vs. time. After an  initial
transient,  family diversity increases  logarithmically in both the TNM and the TAM.
As  cultures merge, several core families end up sharing  the same culture,
and cultural diversity becomes lower than family diversity. Both quantities appear then
to keep growing in a logarithmic fashion.

\begin{figure}
 $
\begin{array}{cc}
\includegraphics[width=0.45\linewidth]{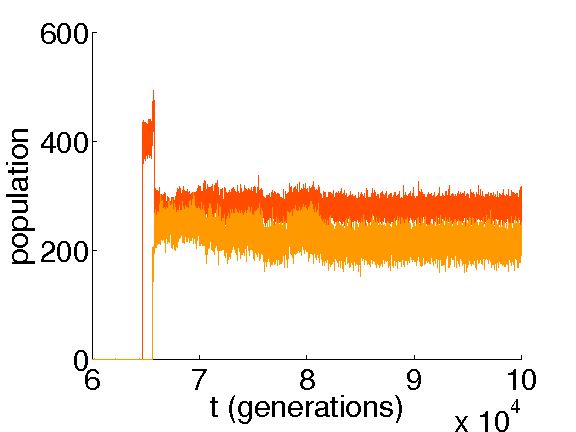}  & 
\includegraphics[width=0.45\linewidth]{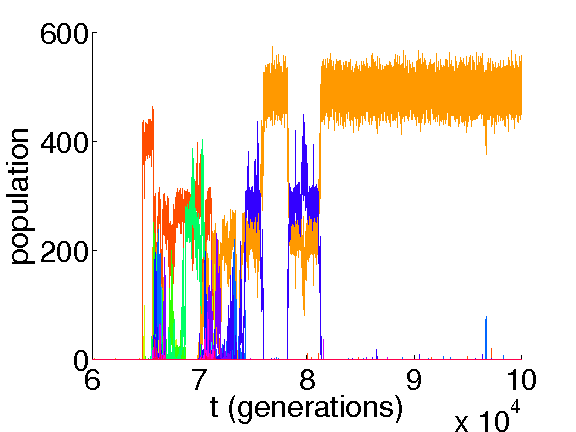}
\end{array}
$ 
\caption{(Color online)
Left: The red and  orange trajectories show the populations of two families (individuals grouped by interaction code) varying   around a 
fixed average during a QESS. Right: The  exact same  individuals   are now grouped by  their culture, each 
culture  depicted in a different color. A number of qualitative changes is seen, where several cultural subgroups arise and finally merge  during the QESS. 
}
\label{Cult} 
\end{figure}

\begin{figure}
 $
\begin{array}{cc}
\includegraphics[width=0.45\linewidth]{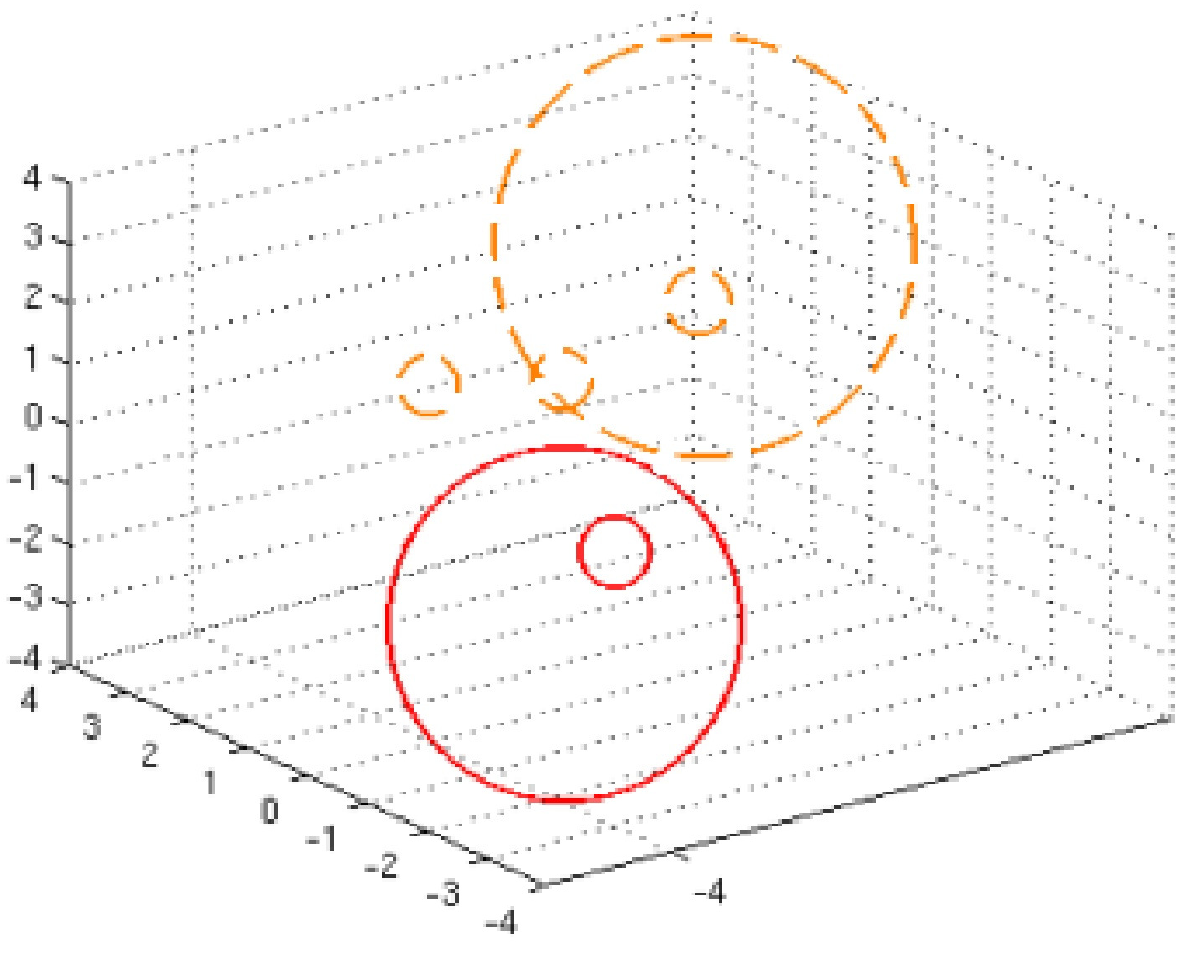}  & 
\includegraphics[width=0.45\linewidth]{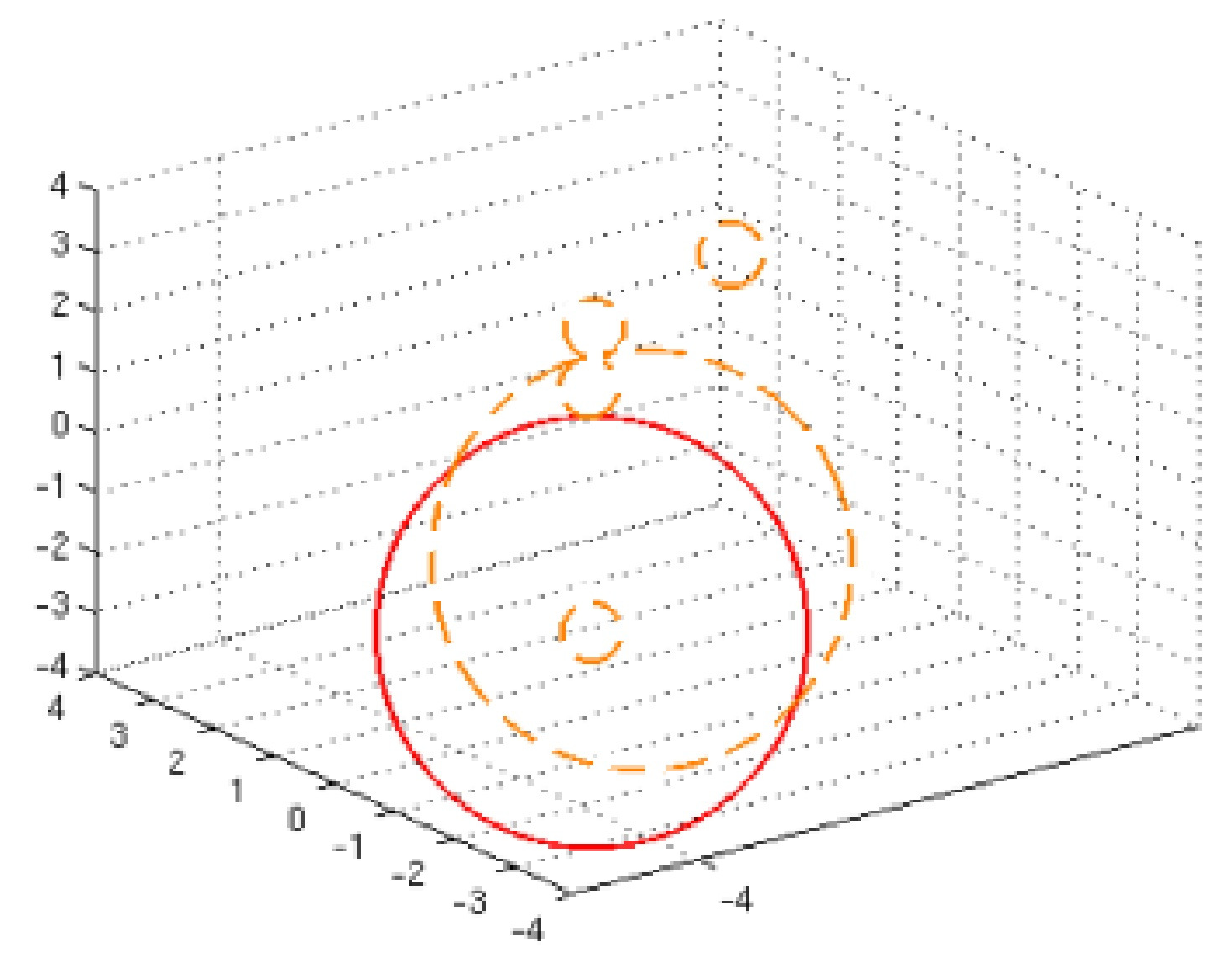}\\
 \includegraphics[width=0.45\linewidth]{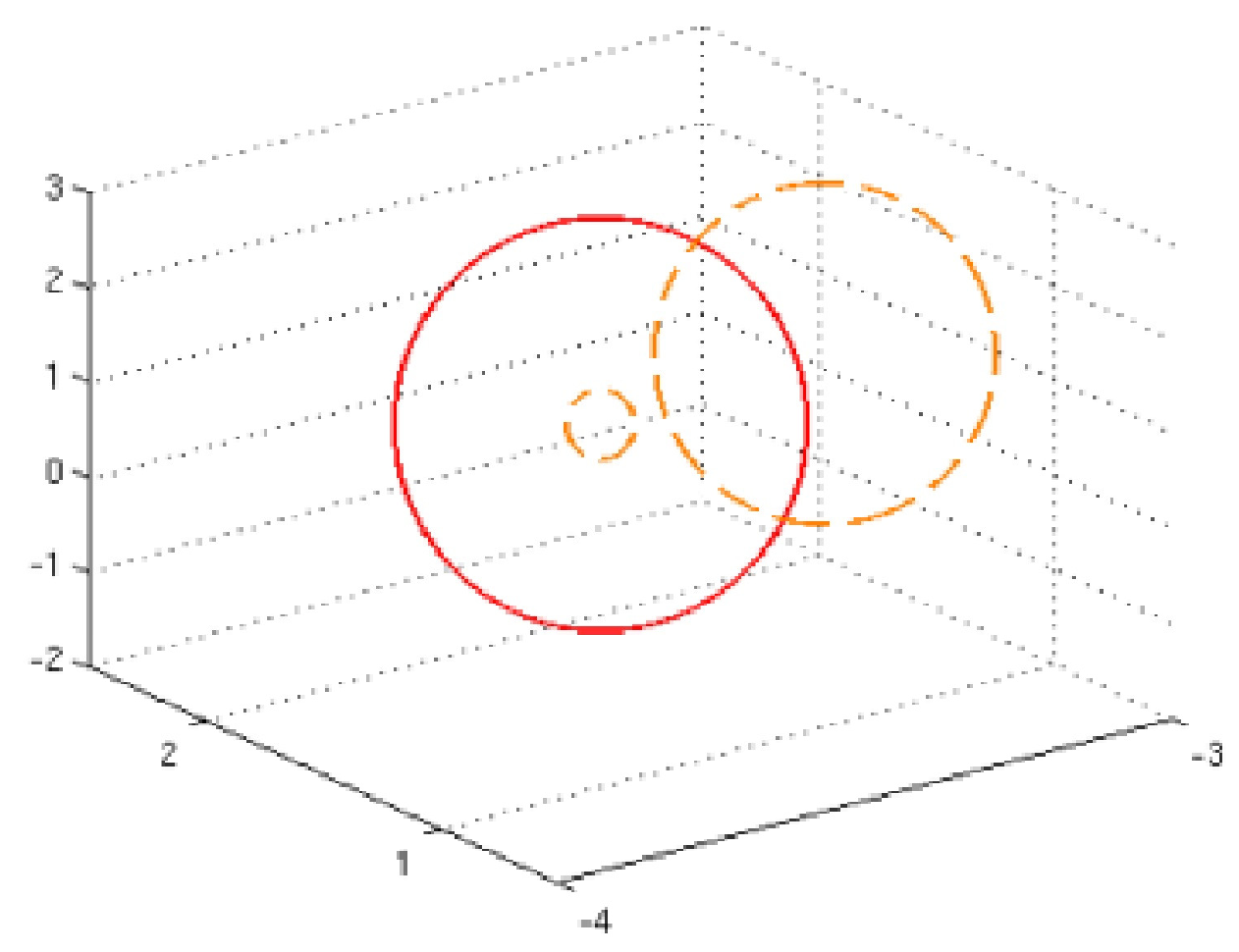} &
\includegraphics[width=0.45\linewidth]{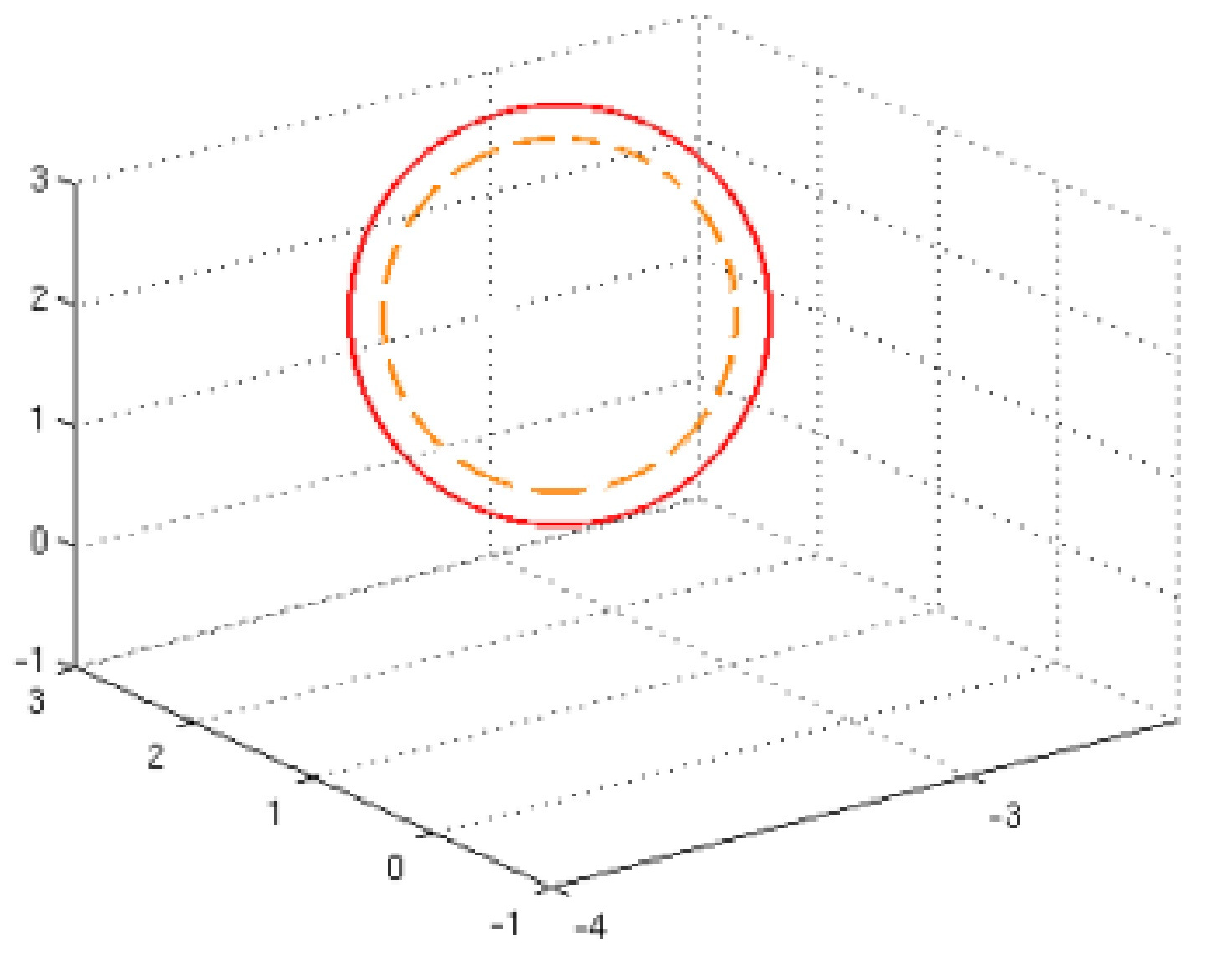} 
\end{array}
$ 
\caption{(Color online) A 3D rendering of the cultural trait exchange between two core families
at times $66\times 10^3, 70\times 10^3, 80\times 10^3$ and $80\times 10^3$ generations, from upper left to lower right.
The center of a circle is the position of the culture, and its area is proportional to the corresponding  population.
Dashed and full lines are used to distinguish between the two cultures.
These plots correspond to the data in Fig.~\ref{Cult}. }
        \label{bubble_plots}
\end{figure}

Figure~\ref{Cult}  shows the time dependence  of two core populations  during a QESS. 
 When agents  are grouped by their interactions, i.e. in  families, 
 two families  are present whose size   exhibits a restricted variation  
around a fixed average.
When  the same individuals are grouped   according to their strategies,  
several cultures are seen to appear, disappear  and  finally merge.
The final merge is possible  in this case,  because the
core populations are both each other's acquaintances and neighbors, i.e. they have non-zero
interactions and are able to copy each other's strategies.
To go  beyond  a macroscopic
description based on  population and  diversity and describe e.g.
 cultural similarities, visualization is called for. 
Using a decimal representation of   
TNM or  TAM   binary strings is not an option,
because  strings differing  by a single bit can have either very different or almost identical
decimal representations, depending on whether the bit in question is the most or least significant one.
To obtain a more faithful 3D representation of our data, we use  Principal Component Analysis (PCA)~\cite{Jolliffe05}, a standard technique 
for data dimension reduction which reasonably maintains distance relationships. 
We start by replacing all zeros in our strings by negative ones, 
whereby  each culture (or  family)  appears as  a point in a twenty dimensional  zero
 centered hypercube, which is naturally embedded in a Euclidean space.
The twenty dimensional cloud of points representing an ecosystem
is  then projected into a 3D cloud, eventually producing Fig.~\ref{bubble_plots}, which illustrates   the cultural
exchanges occurring during a QESS  between  the two extant families
 depicted in  Fig.~\ref{Cult}.

 For convenience and for completeness the steps taken are  summarized below:
We first form  a rectangular matrix $\mathbf Q$,   whose columns are 
vectors of length $20$, each consisting of  a series
of $\pm 1$ and each representing a culture.
The six 
most populous core cultures  are selected as columns of $\mathbf Q$. 
The  square symmetric matrix ${\mathbf O}={\mathbf Q }{\mathbf Q}^T$
has real eigenvalues  and orthogonal eigenvectors, the latter forming  a new basis
for  culture space.
We confine ourselves to  the 3D subspace spanned by 
the eigenvectors corresponding to  the three largest eigenvalues of ${\mathbf O}$,  
and project all our data onto these eigenvectors.
This gives the  3D representation with the largest possible data variation.
To follow   cultural development in time, the eigenvectors   calculated at the 
 `initial' time $t=66000$ generations are   used throughout the analysis.
 A culture's 3D position is  the center of a circle whose area 
  is proportional
 to the corresponding population.
 The color of the circle encodes  the family involved.
Initially, there are two main cultures (the two big circles) flanked by four  less 
populous cultures (the smaller circles). As time goes,  smaller cultures are  gradually absorbed 
by the two larger ones. Even these two eventually merge, forming  a hybrid culture  differing from 
both its predecessors (the two concentric circles).

We now  explore how long time core families survive  in the TAM model and compare  to the 
behavior  of  TNM species and TAM cultures.
In both models, core families and species only disappear through quakes,
while cultures may also disappear more gradually  through   the  TAM copying mechanism.
We might  hence  expect cultures to disappear at a faster rate. This only seems to hold 
at relatively early stages of the evolution process. Old and well entrenched cultures
mainly disappear   together with all their bearers during  quakes and do so at approximately the same rate
as families.
To calculate  the survival probability $S(t_{\rm w})$ of cohorts  of TAM families and cultures 
  and of TNM species, all extant
core families (cultures or species) 
are  counted at times $t_{\rm w}=10^2,10^3,10^4$ and $10^5$. The fraction of the cohorts thus
obtained which still  are 
part of  the core  at time $t>t_{\rm w}$ 
is then logged for $400$ independent 
trajectories. 
The survival probability is finally  estimated by averaging  the fraction remaining over all trajectories.

In the left panel of Fig.~\ref{decay_curves}, the survival probabilities 
for TAM cultures (red) and families  (green) are plotted 
on log-log scales together with those of TNM species (blue).
As mentioned, at early times, cultures decay faster than both 
families and species.
This means that, after a while,  several families   share the same
culture, lading to  fewer starting point for mutations. This stunts the exploration
 of configuration space and makes the core highly stable. 
At these later times all curves approximately decay
at the same rate.

A sufficiently large core family or species disappears, usually  together with the rest of the core, when a quake hits,
while   small core species might  drift away from the core due to a decrease in their population.
Neglecting the last possibility,
 the probability of family (or species) survival  from $t_{\rm w}$ to $t$  is  the same as the  probability that no 
quake hits in that time interval. Since quakes are approximately log-Poisson distributed~\cite{Sibani03}
we expect 
\begin{equation}
S_{\rm family}(t_{\rm w}) \approx  \left( \frac{t}{t_{\rm w}}\right)^{-x},
\label{survival}
\end{equation} 
for some exponent $x>0$. The same reasoning applies to TNM species.
In  the right panel on Fig.~\ref{decay_curves}, all three survival probabilities
are on log-log scales plotted vs. $\frac{t}{t_{\rm w}}$. The TNM species survival data
(blue curves) fall in two groups, with the two data sets collected at late times and the two collected at early times 
nearly overlapping. The same applies to the TAM family survival probabilities (green).
The TAM culture survival probabilities only overlap at late times, in agreement with our previous remarks.

Clearly, $\frac{t}{t_{\rm w}}$ scaling  holds approximatively for  the data
collected at late times, i.e. for sufficiently large $t_{\rm w}$. 
The exponent $x$ for the power-law decay of the survival probability in that region   is
estimated as  $x\approx 0.2$. Hence the  probability density function for a family, species or
culture lifetime  
\begin{equation}
P(t_{\rm w}) = -\frac{d}{dt}  S(t_{\rm w}) =  x t_{\rm w}^{-x-1} \left( \frac{t}{t_{\rm w}}\right)^{-x-1}.
\label{survivalPDF}
\end{equation} 
Even though the value of $x$ is uncertain, a finite average life-time for families, species or cultures would
require $x>1$, which can safely be excluded.
\begin{figure}
 $
\begin{array}{cc}
	\includegraphics[width=0.45\linewidth]{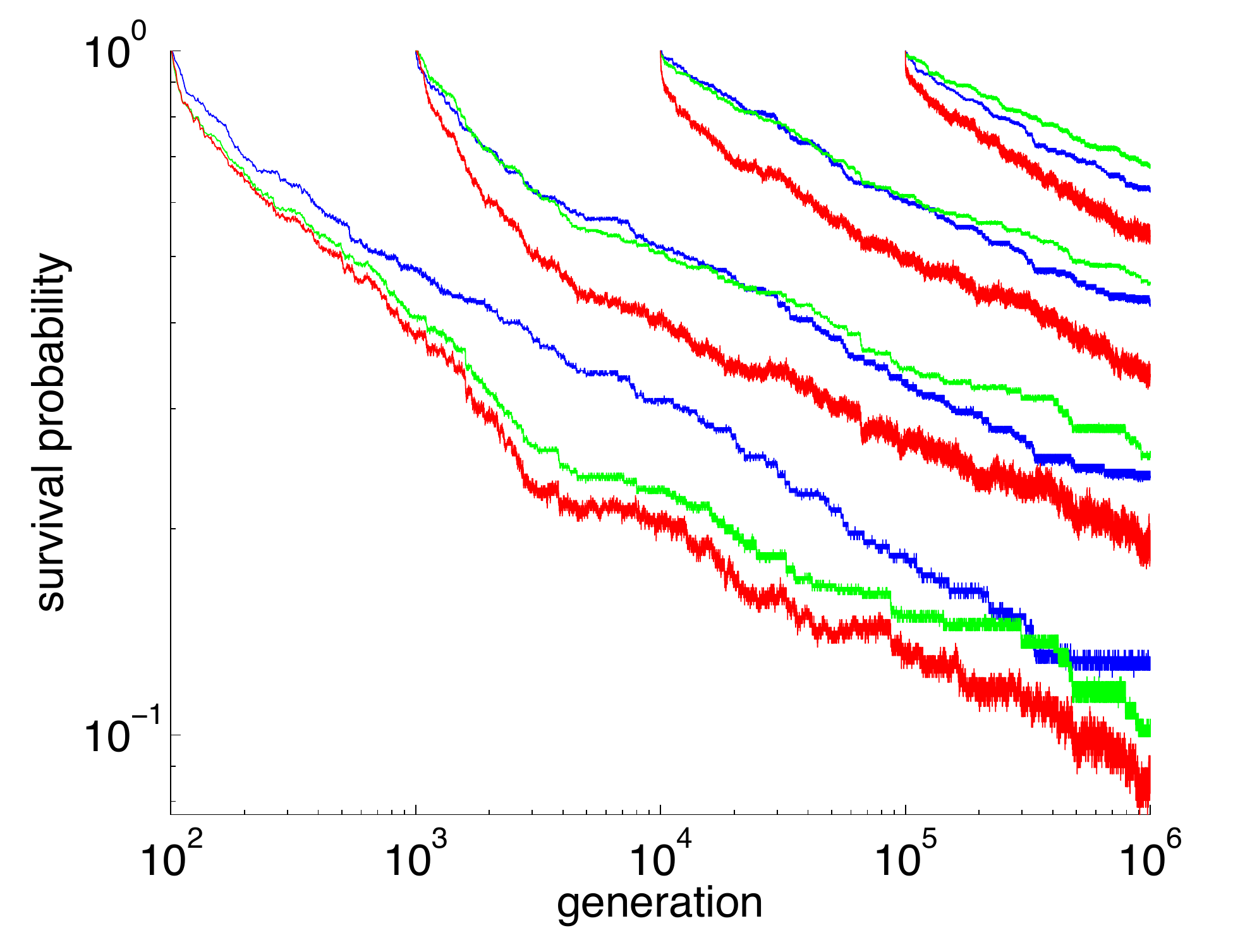}  & 
\includegraphics[width=0.45\linewidth]{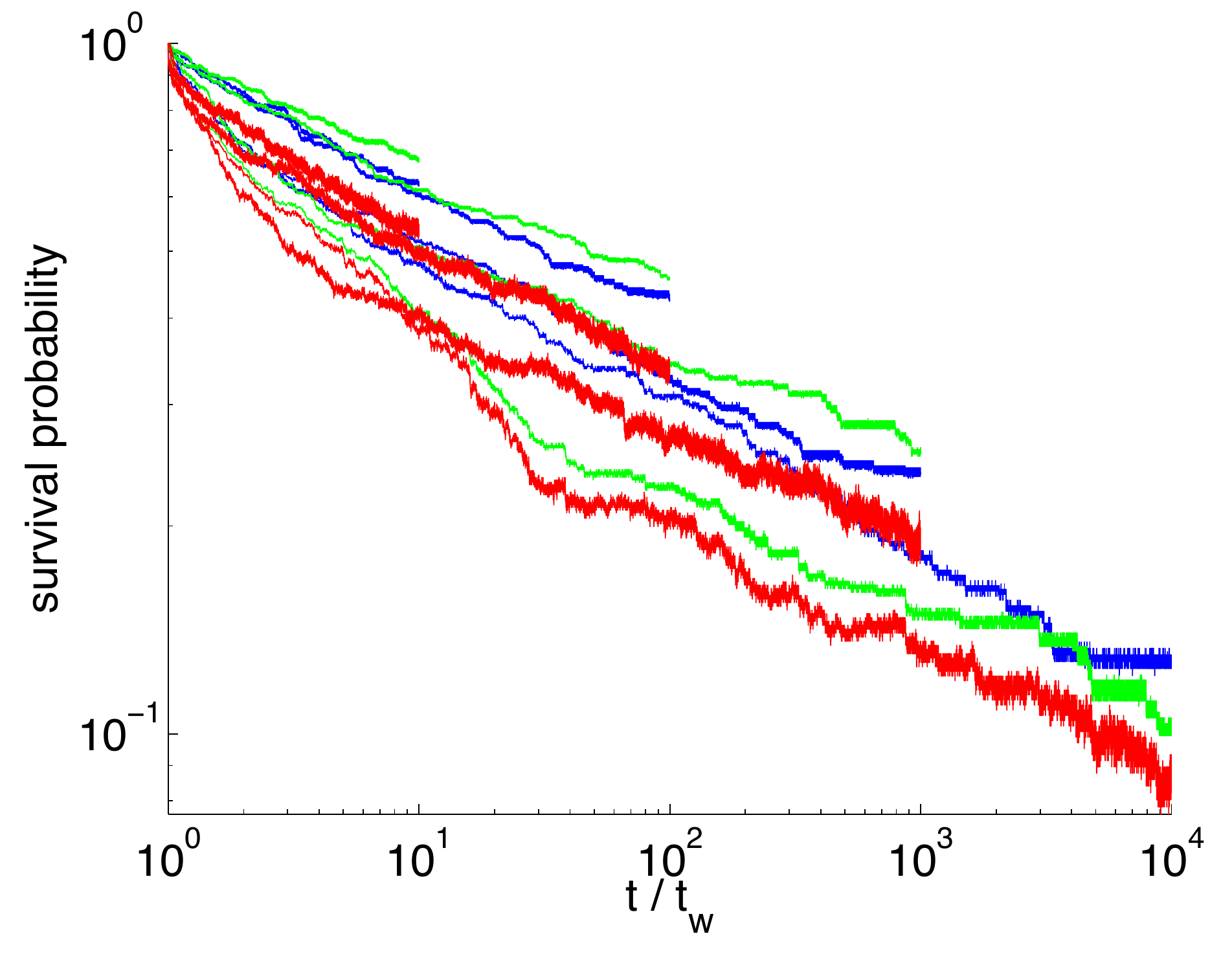}
\end{array}
$ 
\caption{ Left: The survival probability of a cohort present at $t_{w}$ 
and consisting of TAM families (green), TNM species (blue)
and  TAM cultures (red).
  Right: Same  data, now  plotted versus $t/t_{w}$.
	}
 \label{decay_curves}
\end{figure} 
The lack of a finite average life time implies that  empirically collected life-times 
would have a large scatter.

In the early stages of a QESS  establishment, i.e. soon after a quake, the TAM always  produces a flurry
of short lived cultures, which eventually disappear. Intuitively, a similar
situation could  be expected  in human cultural setting, soon  after 
a new disruptive technology enters the scene. Recent examples could be the introduction of personal computers
and, later, of cell phones. 
\begin{figure}
 $
\begin{array}{cc}
	\includegraphics[width=0.45\linewidth]{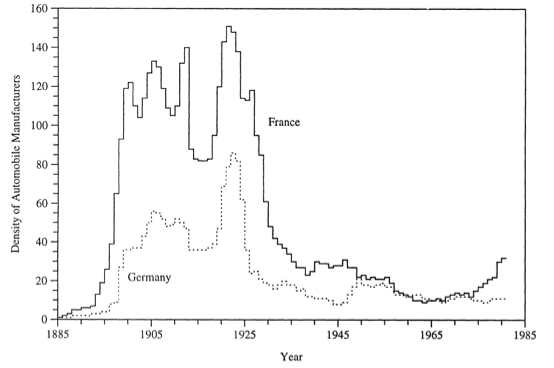}  & 
\includegraphics[width=0.45\linewidth]{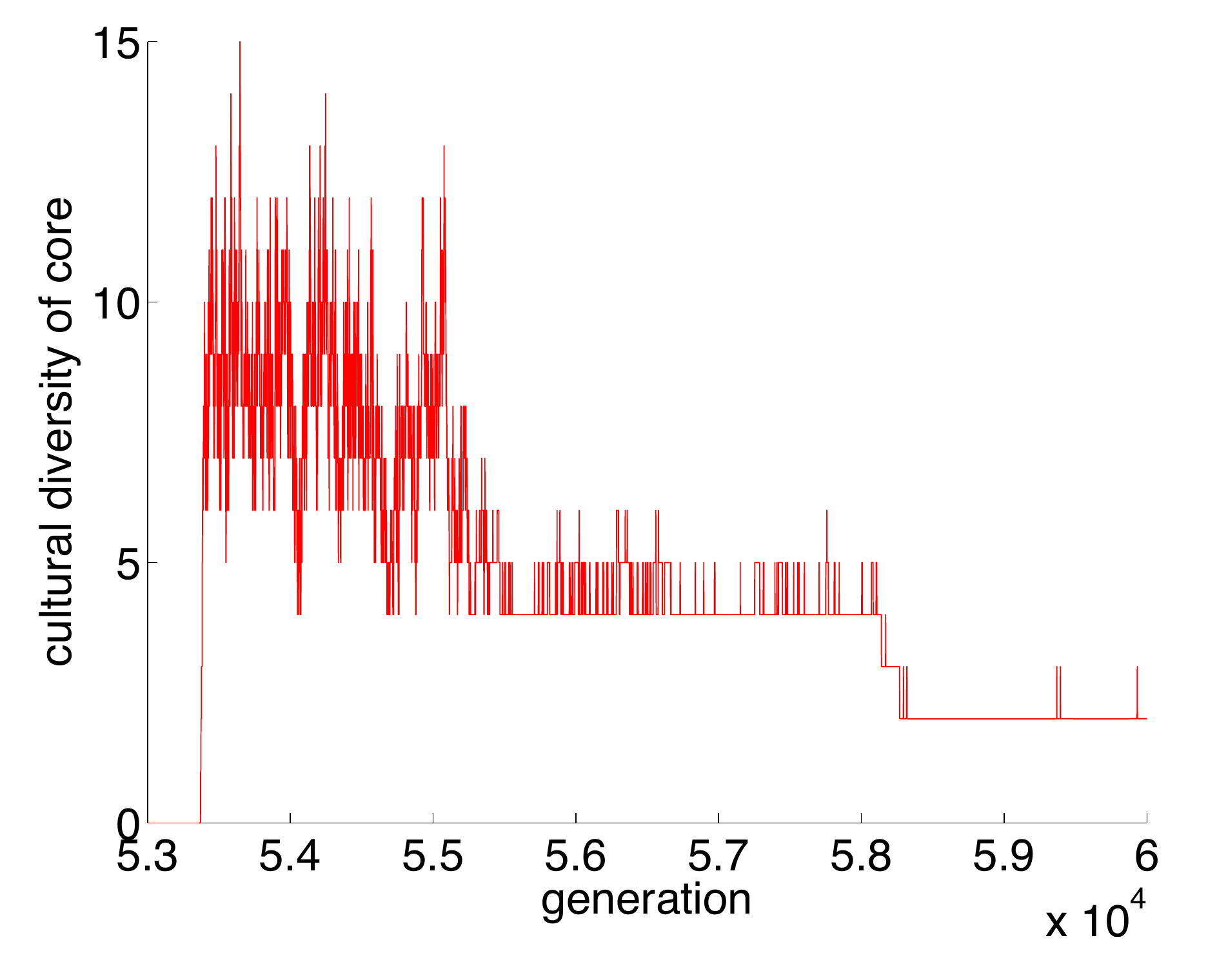}
\end{array}
$ 
\caption{ Left: The number of car manufacturers  in  Germany and France.
(Data reproduced from Ref.~\cite{Hannan95}). 
Right: The number of cultures in the TAM immediately after a quake.
	}
 \label{decay_curves}
\end{figure} 
The example we will discuss is older, and concerns 
the number of firms in the automobile manufacturing industry from 1886 to 1981, i.e. in a period
starting soon after the automobile was introduced to the market. We consider `ways to build automobiles'
to be attributes of car companies which are  somewhat similar to TAM cultures. The parallel is admittedly
incomplete, since \emph{i)} car companies are all different, and the equivalent of a population sharing a culture is unclear,
and \emph{ii)} the interactions of car companies with each other, with their suppliers and with their customers
is not given.
Our---thus    merely qualitative---comparison utilizes   data   stemming  from
 Ref.~\cite{Hannan95}, where firms were counted that declared intentions to manufacture automobiles for the market.
 The birth and  death of a firm are  the dates when production commences, respectively ends.
 The left panel of Fig.~\ref{decay_curves} shows the 
 number of car manufacturers in Germany and France vs. time, while the right panel
 shows  the number of different cultures in a single trajectory of the TAM right after a quake.
 After an initial slow start, the number of automobile manufacturers rapidly increases to a plateau
 that lasts a few decades. The great variation seen during this period matches
 the great scatter expected for company lifetimes.
 Eventually, the number of companies ebbs to a much lower level  of `entrenched' manufacturers,
 those which, based on  TAM properties, would only cease to build cars once   the automobile itself is
 supplanted by a different product. The TAM data are rather similar:  due to the flurry of activity right after a quake,
 the number of cultures has a broad peak and then slowly tapers off.

\section{Discussion and outlook}

The TAM  dynamics is heavily based on the TNM model of
 biological evolution, and leads to the formation of a 
multicultural and evolving ecosystem, where long periods of stability labeled
by extant families  replace each other through rapid quakes,  similarly  to the  QESS
of the TNM.
 In a cultural setting, quakes could correspond to
innovations disrupting the existing know-how and  radically changing the way societies are 
organized.  The hectic cultural exchanges   accompanying these  quakes
are  also reminiscent of aspects of cultural evolution.
TAM cultures get more stable with age and  lack a finite average life-time, leading to a large and time increasing
scatter in the empirical distribution of cultures. 
 At least qualitatively, these features resemble some aspects of  real
 cultural evolution. Yet, TAM agents follow  dynamical
rules with  a high degree of randomness and  have  limited
 information on  the situation of other agents.

The only real difference  between TAM and TNM dynamics
lies in the copying and mutation mechanism: The TAM `genome' is divided into two parts, 
 called `interaction genome'
and `strategy'.  Random mutations can only hit the second part directly. In a biologically inspired  interpretation  this 
could mean that  the mutations affecting the first part are  never  viable. Secondly, individuals can 
copy each other's strategies or parts thereof, with the choice of what to copy biased by the  frequency
or `popularity' of the strategies copied.
This  swapping is similar to genetic
recombination in bacteria. Note however that the new genetic material incorporated 
by an individual has no \emph{immediate}
effect on the latter's  reproductive ability. The effect comes first when  an  intervening mutation
 `promotes'  the copied and mutated material into the 
`interaction genome' of the individual. 

The  macroscopic dynamics of both  the TAM and the TNM and that 
of  biological macroevolution~\cite{Sibani95,Newman99e} are   decelerating, while  it is  commonly 
believed that human cultural evolution is an accelerating process. This point of view was
recently  challenged in a comparative study of cultural and biological
evolution rates~\cite{Perreault12}, where both types of rates are found to decrease with the 
inverse  of the observation time over which they are measured.
This  behavior   might be consistent with the deceleration of the TNM and  TAM 
dynamics.

 The perceived acceleration of 
  cultural evolution might,  at least in part,
be due to  time being  measured
in physical units. These units are appropriate for biological
evolution,  where  the rate of mutation events (successful or not) can be considered
constant in time. The same units are not necessarily appropriate,  we would claim, for human cultural evolution,
especially during periods where the population varies strongly.
The  intensity of inter-human interactions, i.e.
their  number  per unit of (physical) time,  
has risen thousandfold  from, say,  neolithic times, where population was sparse and
communication slow,  to present times.
 The intensity   has risen at an even   faster pace during the last  century, 
due to the increased levels of urbanization and, lately,  to our pervasive and fast communication
networks. 
It   stands to reason that   human cultural evolution 
should depend on  the number of  inter-personal  interactions,
and that time should be rescaled  to units where 
the rate of these interactions is constant.
Such  rescaling  would be very difficult to carry out, but it certainly   inflate our current time compared to, say, neolithic time.
Hence,  the modern pace of human evolution expressed in rescaled    units would come out greatly reduced.

Let us finally note they the population increase humanity has experienced 
is closely associated to emerging  technologies and to the new cultural 
settings which allow their exploitation~\cite{Leland00}.
In the current version of the TAM
cultures already  affect,   via the copying/mutation
mechanisms, the  
agents' reproduction rates, but the coupling is random.

In summary, the agents of the TAM model act in a probabilistic way based on partial knowledge
of their environment. The emergent properties of the model are its core, a group of established 
and  mutually supportive cultures which could qualitatively correspond to successful companies
trading with each other, and its  intermittently populated cloud,  similar to the large number 
of start-ups which quickly go bankrupt every year. At least in the model, disruptive
innovations originate from `destabilizer' start-ups, which grow and perish but trigger wave of change 
eventually leading to new organizational ecosystems.
\\
\noindent \emph{Acknowledgments.}
The authors are indebted to Guido Fioretti for his insightful comments and for providing 
an extensive and commented  bibliography. Rudy Arthur has helped with  his 
interest and support through the different stages of the work, and 
Nikolaj Becker has shared his  insights on the dynamics of the TNM.
\bibliographystyle{unsrt}
\bibliography{paolo}
 \end{document}